\documentclass[twocolumn,showpacs]{revtex4}
\usepackage{dcolumn}
\usepackage{bm}
\usepackage{amsmath}
\usepackage{amssymb}
\usepackage{amsfonts}
\usepackage{graphicx}

\begin{document}

\title{On the negative spectrum of the $2+1$ black hole}
\author{Olivera Mi\v{s}kovi\'{c} }
\email{olivera.miskovic@ucv_dot_cl}
\affiliation{Instituto de F\'{\i}sica, P. Universidad Cat\'{o}lica de Valpara\'{\i}so,
Casilla 4059, Valpara\'{\i}so, Chile.}
\author{Jorge Zanelli}
\email{z@cecs_dot_cl}
\affiliation{Centro de Estudios Cient\'{\i}ficos (CECS), Casilla 1469, Valdivia, Chile.}
\date{\today }

\begin{abstract}
In (2+1)-dimensional gravity with negative cosmological constant, the
states\ in the negative energy range, between AdS ($M=-1$) and the so-called
BTZ black hole ($M\geq 0$), correspond to topological defects with angular
deficit $0<\alpha <2\pi $. These defects are produced by (static or
spinning) $0$-branes which, in the extreme case $M\ell =-|J|$, admit
globally-defined covariantly constant spinors. Thus, these branes correspond
to BPS solitons and are stable ground state candidates for the corresponding
supersymmetric extension of 2+1 AdS gravity. These branes constitute
external currents that couple in a gauge-invariant way to three-dimensional
AdS gravity.
\end{abstract}

\maketitle

\section{Introduction}

Gravity in 2+1 dimensions with negative cosmological constant is possibly
the simplest realistic analogue of\ General Relativity. In spite of having
no Newtonian attraction, it gives rise to nontrivial black hole solutions
that in many ways resemble astronomic black hole candidates observed at many
galactic nuclei \cite{BTZ}. For a comprehensive review, see \cite{Carlip}.
In an appropriate coordinate system, the metric%
\begin{equation}
ds^{2}=-f^{2}d\tau ^{2}+\frac{d\rho ^{2}}{f^{2}}+\rho ^{2}\left( Nd\tau
+d\phi \right) ^{2},  \label{BTZ metric}
\end{equation}%
where $f^{2}=-M+\frac{\rho ^{2}}{\ell ^{2}}+\frac{J^{2}}{4\rho ^{2}}$ and $%
N=-\frac{J}{2\rho ^{2}}$, describes a black hole with mass $M$ and angular
momentum $J$, provided $M\geq |J|\ell ^{-1}\geq 0$. For $M=-1$ (and $J=0$)
the spacetime is globally AdS and has no singularity. For the intermediate
range $0>M>-1$, there is no horizon surrounding the singularity at the
origin and therefore these solutions are naked singularities (\textbf{NS}).

The purpose of this letter is to discuss the nature of the NS represented by
the gap in the spectrum separating the real black holes ($M\geq 0$) from the
anti-de Sitter \textquotedblleft vacuum\textquotedblright\ ($M=-1$). Our
observation is that these NS states are not completely unphysical, but
correspond to topological defects produced by a $0$-brane at $r=0$. Black
holes ($M\geq 0$) on the other hand, are not produced by matter sources
coupled to gravity: there is nothing at $r=0$. The black hole is a purely
gravitational configuration without a matter source curving spacetime around
it.

The $0$-branes couple to gravity in a gauge-invariant way, a feature that
might be instrumental in setting up a perturbative scheme for quantization
and that can be generalized to higher dimensions. Moreover, these NS can
also have nonvanishing angular momentum and, in the extreme case $M=-|J|\ell
^{-1}$, the geometry admits one globally defined Killing spinor. These
extremal $0$-branes behave as solitons that saturate the Bogomolny'i bound
--BPS states--, and are therefore possible stable vacua for 2+1 supergravity.

According to our present understanding of gravity, singularities certainly
form in gravitational collapse. What is not so certain is whether event
horizons that protect outside observers from the singularity necessarily
form as well. Since NS can break predictability and raise a number of
conceptual puzzles \cite{Earman}, it would be comforting if Penrose's cosmic
censorship hypothesis were a theorem in classical GR. Numerical experiments,
however, show that in a wide variety of collapse scenarios a horizon may not
form at all, leaving a singularity exposed to\ an outside observer \cite%
{NumericalExp}. The nature of these singularities and their potential as
laboratories where quantum gravity effects could be studied, make them
interesting objects of analysis.

Topological defects are examples of rather harmless NS produced by
identification with a Killing vector field that leaves fixed points in a
manifold. The singularity is a submanifold that concentrates a deficit
angle, like the apex of a cone in a two-dimensional Euclidean plane with an
angular identification \cite{Jackiw90}. A cone can be described locally as a
plane in coordinates $\left( x^{1},x^{2}\right) =\left( r\cos \phi
_{12},r\sin \phi _{12}\right) $, where the radial coordinate takes values $%
0\leq r<\infty $, and the azimuthal angle $\phi _{12}$ has a deficit, $0\leq
\phi _{12}\leq 2\pi (1-\alpha )$. This topological defect is a naked
singularity at the apex of a cone, $r=0$, where the curvature is infinite,
described by a $\delta $-function.

In these coordinates the metric has the standard flat form, $%
ds^{2}=dr^{2}+r^{2}d\phi _{12}^{2}$, but the identification $\phi
_{12}\simeq \phi _{12}+2\pi (1-\alpha )$, with $0\leq \alpha <1$, produces
the singularity at $r=0$. A standard azimuthal angle $\phi $ of period $2\pi
$ can be introduced by rescaling $\phi _{12}=\left( 1-\alpha \right) \phi $,
making manifest the topological defect through a factor multiplying the
angular sector of the metric. The resulting Riemann curvature two-form in
two dimensions is $R_{\text{\ \ }2}^{1}=d\omega _{\text{\ \ }2}^{1}$, where $%
d\omega _{\text{\ \ }2}^{1}=-d\phi _{12}$. Then, as pointed out in \cite%
{Jackiw90}, the identity $dd\phi _{12}=-2\pi \alpha \,\delta
(T_{12})\,d\Omega _{12}$ is valid in the sense of Stokes' theorem upon
integration, and a curvature singularity
\begin{equation}
R_{\text{\ \ }2}^{1}=2\pi \alpha \,\delta (T_{12})\,d\Omega _{12}
\end{equation}%
is found at the origin, where $\delta (T_{12})d\Omega _{12}$ is the Dirac
delta two-form with support at $r=0$ on the two-dimensional plane $T_{12}$\
(in polar coordinates). It can also be checked that the torsion tensor
vanishes, thanks to the property of the Dirac distribution, $r\delta (r)=0$.
The result can be re-interpreted in terms of the identification by the
Killing vector for rotational symmetry around the origin in the 1-2 plane, $%
\partial _{\phi _{12}}=x_{1}\partial _{2}-x_{2}\partial _{1}$. The angular
defect results from the identification $x^{i}\simeq x^{i}+\xi ^{i}$,%
\begin{equation}
\left(
\begin{array}{c}
x^{1} \\
x^{2}%
\end{array}%
\right) \simeq \left(
\begin{array}{cc}
\cos 2\pi \alpha & \sin 2\pi \alpha \\
-\sin 2\pi \alpha & \cos 2\pi \alpha%
\end{array}%
\right) \left(
\begin{array}{c}
x^{1} \\
x^{2}%
\end{array}%
\right) \,,  \label{Xc,Yc}
\end{equation}%
produced by the Killing vector field $\xi =\xi ^{i}\partial _{i}=-2\pi
\alpha \,\partial _{\phi _{12}}$. Note that the strength of the curvature
singularity in the identified geometry equals the magnitude of the angular
deficit, $2\pi \alpha $, times the two-form delta distribution, $\delta
(T_{12})d\Omega _{12}$, which can be identified as the source of the conical
singularity. This happens whenever a curvature singularity is the result of
an identification by a spacelike Killing vector that leaves fixed points
\cite{Miskovic-Zanelli09}.

\section{Defects in 2+1 AdS gravity}

Three-dimensional gravity with negative cosmological constant can be
described by the CS Lagrangian for the $so(2,2)$ algebra, with connection
\begin{equation}
A=\frac{1}{2}\,\omega ^{ab}J_{ab}+\frac{1}{\ell }\,e^{a}J_{a}\,,
\end{equation}%
where $J_{ab}$ and $J_{a}$ are the generators of Lorentz rotations and AdS
boosts. The corresponding AdS curvature is $F=\frac{1}{2}\,\left( R^{ab}+%
\frac{1}{\ell ^{2}}\,e^{a}e^{b}\right) J_{ab}+\frac{1}{\ell }\,T^{a}J_{a}$,
where $R_{\text{\ \ }b}^{a}=d\omega _{\text{\ \ }b}^{a}+\omega _{\text{\ \ }%
c}^{a}\omega _{\text{\ \ }b}^{c}$ and $T^{a}\equiv De^{a}=de^{a}+\omega _{%
\text{ \ }b}^{a}\,e^{b}$ are the Riemann curvature and the torsion $2$%
-forms, respectively.

A topological defect, analogous to the conical singularity described above,
can also be produced in the global AdS geometry, through an identification
by the Killing vector field $\xi =-2\pi \alpha \,\left( x_{1}\partial
_{2}-x_{2}\partial _{1}\right) $, that leaves invariant the 1-2 plane of the
three-dimensional AdS spacetime. The fixed points of the Killing field
become the support of a source. The resulting geometry has an angular defect
of magnitude $2\pi \alpha $ and the metric reads \cite{Miskovic-Zanelli09}%
\begin{equation}
ds^{2}=-\left( \frac{r^{2}}{\ell ^{2}}+1\right) \,dt^{2}+\frac{dr^{2}}{\frac{%
r^{2}}{\ell ^{2}}+1}+(1-\alpha )^{2}r^{2}d\phi ^{2}.
\label{AdS+point-source}
\end{equation}%
Here $0\leq \phi \leq 2\pi $ is periodic; for $\alpha =0$ the conical
singularity at $r=0$ disappears and the geometry becomes globally AdS.
Direct computation confirms the curvature singularity at the origin of the $%
(x^{1},x^{2})=(r\cos \phi _{12}\,,r\sin \phi _{12})$ plane, and the AdS
curvature is
\begin{equation}
R^{ab}+\frac{1}{\ell ^{2}}\,e^{a}e^{b}=2\pi \alpha \,\delta (T_{12})d\Omega
_{12}\,J_{12}\,\eta ^{\lbrack 12][ab]}\,,  \label{AdS-curvature}
\end{equation}
where $\eta ^{\lbrack 12][ab]}=\eta ^{1a}\eta ^{2b}-\eta ^{1b}\eta ^{2a}$,
and the torsion vanishes. From the Chern-Simons field equations, $F=j$, the $%
0$-brane source at the defect can be read from the right hand side of Eq.(%
\ref{AdS-curvature}),%
\begin{equation}
j(x)=2\pi \alpha \,\delta \left( T_{12}\right) d\Omega _{12}\,J_{12}\,.
\label{AdS-j-0}
\end{equation}

Note that the norm of the Killing vector $\left\Vert \xi \right\Vert
^{2}=4\pi ^{2}\alpha ^{2}r^{2}$ is positive for $r\neq 0$. Hence, the
identification takes place in the Euclidean ($x^{1},x^{2}$)-plane, and
vanishes at the singularity. The time-like Killing vector $\partial _{t}=-%
\frac{1}{\ell }$ $\left( x_{0}\partial _{3}-x_{3}\partial _{0}\right) $,
that commutes with $\xi $, is everywhere timelike, $\left\Vert \partial
_{t}\right\Vert ^{2}=-\left( 1+\frac{r^{2}}{\ell ^{2}}\right) <0$.

This point source at rest at the origin of the spatial section looks
suspiciously similar to a black hole: it is a localized, static, spherically
symmetric, locally AdS geometry. Indeed, one can write the metric (\ref%
{AdS+point-source}) in Schwartzschild-like coordinates by rescaling $\rho
=r\left( 1-\alpha \right) $ and $\tau =\frac{t}{1-\alpha }$, and
\begin{equation}
ds^{2}=-\left( \frac{\rho ^{2}}{\ell ^{2}}-M\right) d\tau ^{2}+\frac{d\rho
^{2}}{\frac{\rho ^{2}}{\ell ^{2}}-M}+\rho ^{2}d\phi ^{2},
\label{Negative-M-B-H}
\end{equation}%
which looks like the black hole in $2+1$ dimensions \cite{BTZ}. However,
this is only an illusion because here the \textquotedblleft
mass\textquotedblright\ is negative: $M=-(1-\alpha )^{2}$, which shows that
this solution is a naked singularity. This is correct, since branes are
accessible (naked) singularities in spacetime and not protected by a
horizon. In contrast with this $0$-brane, the black hole ($M>0$) results
from an identification with a Killing vector, that does not have fixed
points in the embedding space, $\mathbb{R}^{2,2}$ \cite{BHTZ}. The BTZ
manifold is topologically a cylinder and has no conical singularities, while
the spacetime (\ref{AdS+point-source}) is an orbifold.

The geometry described above is static, but a spinning $0$-brane can also be
obtained by identification with an appropriate Killing vector. The resulting
stationary, axisymmetric spacetime is also described by (\ref{BTZ metric})
where\textbf{\ }$f^{2}$ and $N$ have the same form as for the black hole but
with $M<0$, and is again a naked singularity. This geometry can be seen as
the result of an identification generated by two globally defined,
independent Euclidean rotations in the embedding space. Consider the
following parametrization of a pseudosphere in $\mathbb{R}^{2,2}$,%
\begin{equation}
\begin{array}{ll}
x^{0}=A(\rho )\,\cos \phi _{03}\,,\qquad & x^{1}=B(\rho )\,\cos \phi _{12}\,,
\\
x^{3}=A(\rho )\,\sin \phi _{03}\,, & x^{2}=B(\rho )\,\sin \phi _{12}\,,%
\end{array}
\label{spinning}
\end{equation}%
where $A^{2}-B^{2}=-\ell ^{2}$ with $A$ and $B$ chosen as $A=\sqrt{(\rho
^{2}+\ell ^{2}a^{2})/(a^{2}-b^{2})}$ and $B=\sqrt{(\rho ^{2}+\ell
^{2}b^{2})/(a^{2}-b^{2})}$. If the angles in the 0-3 and 1-2 planes have the
form $\phi _{03}=b\phi +\frac{a\tau }{\ell }$ and $\phi _{12}=a\phi +\frac{%
b\tau }{\ell }$, the (real) parameters $a$ and $b$ can be related to the
mass ($M\leq 0$) and angular momentum through
\begin{equation}
a\pm b=\sqrt{-M\pm \frac{J}{\ell }\,}.
\end{equation}
Thus, $a^{2}-b^{2}=\sqrt{M^{2}-\frac{J^{2}}{\ell ^{2}}}\geq 0$ provided $%
0<\left\vert J\right\vert \leq \ell \left\vert M\right\vert $, and the
static $0$-brane is recovered for $b=0$.

The spinning $0$-brane results from a single identification in AdS space, $%
\phi \simeq \phi +2\pi $, but it can also be seen as produced by two
independent identifications (in different planes) in $\mathbb{R}^{2,2}$, $%
\phi _{12}\simeq \phi _{12}+2\pi a\,$and $\phi _{03}\simeq \phi _{03}+2\pi b$%
, so that the corresponding angular deficits are $\alpha =1-a$ and $b$ \cite%
{Freedom}. The Killing vector $\xi $ that produces the identification $%
x^{A}\simeq e^{\xi }\,x^{A}$, turns out to be a linear combination of two
independent isometries, $\xi =-2\pi \alpha \,J_{12}+2\pi b\,J_{03}$ ($%
J_{AB}=x_{A}\partial _{B}-x_{B}\partial _{A}$). In $\mathbb{R}^{2,2}$, the
identifications along $J_{12}$ and $J_{03}$ lead to two conical
singularities in the corresponding planes,
\begin{equation}
dd\phi _{12}=-2\pi \alpha \,\delta (T_{12})\,d\Omega _{12}\,,\;dd\phi
_{03}=2\pi b\,\delta (T_{03})\,d\Omega _{03}\,.  \label{dd}
\end{equation}%
In the covering AdS space, defined by the pseudosphere $x\cdot x=-\ell ^{2}$%
, these identifications correspond to the single identification $\phi \simeq
\phi +2\pi $ which, in turn, implies\ $dd\phi \neq 0$.

The AdS curvature can be calculated working directly with the metric%
\begin{equation}
ds^{2}=\left( B^{\prime 2}-A^{\prime 2}\right) \,d\rho ^{2}-A^{2}d\phi
_{03}^{2}+B^{2}d\phi _{12}^{2}\,.  \label{spinning metric}
\end{equation}%
The result is that the curvature has the form of the field equations $F=j$,
where using the topological identities (\ref{dd}), one can read the source
of this spinning $0$-brane as
\begin{equation}
j=2\pi b\,G_{03}\,\delta (T_{03})\,d\Omega _{03}+2\pi \alpha
\,G_{12}\,\delta (T_{12})\,d\Omega _{12}\,,  \label{spinning j}
\end{equation}%
with the two commuting generators
\begin{equation}
G_{03}=\frac{a\,J_{03}+b\,J_{01}}{\sqrt{a^{2}-b^{2}}}\,,\qquad G_{12}=\frac{%
a\,J_{12}-b\,J_{23}}{\sqrt{a^{2}-b^{2}}}\,.
\end{equation}%
Note that this form of the current corresponds to two mutually commuting
independent $U(1)$ sources, corresponding to the Cartan subalgebra of the
rank 2 AdS group $SO(2,2)$. In the limit $J=0$ (or $b=0$), the above
expression reduces to the previous result, $j=2\pi \alpha \,\delta
(T_{12})\,d\Omega _{12}J_{12}$.

The extremal spinning $0$-brane has to be addressed separately because the
transformation (\ref{spinning}) is not defined when $a=b$ (that is, for $%
\left\vert M\right\vert \ell =\left\vert J\right\vert $). However, the
Killing vector $\xi $ has a well-defined extremal limit given by
\begin{equation}
\xi \equiv \frac{1}{2}\,\xi ^{AB}J_{AB}\rightarrow 2\pi \alpha \,\left(
J_{03}-J_{12}\right) -2\pi J_{03}\,,
\end{equation}%
where the last term represents a rotation by $2\pi $ and can be omitted.(see
\cite{Freedom})

From the point of view of the embedding flat space, two independent
rotations in Euclidean planes given by two independent generators $J_{03}$
and $J_{12}$ combine into only one generator, $J_{03}-J_{12}$. Thus, the
Killng vector changes its character, which can be seen from the
corresponding Casimir invariants $I_{1}=\xi ^{AB}\,\xi _{AB}$ and $%
I_{2}=\varepsilon ^{ABCD}\,\xi _{AB}\,\xi _{CD}$, that are also continuous
in this limit, but correspond to a different type in the classification
given in Ref. \cite{BHTZ}. From the continuity of the invariants, one can
expect that there exists a Killing vector whose identification introduces a
topological defect in the flat space that produces an extremal $0$-brane.
The explicit form of this Killing vector can also be found from the
identification needed to turn the pseudosphere $\eta _{AB}\,x^{A}x^{B}=-\ell
^{2}$ into the extreme black hole metric (\ref{BTZ metric}) with $J=-\gamma
\ell M$, where $\gamma =\pm 1$, so that the lapse and shift functions in the
metric take the form $f=\frac{\rho }{\ell }-\frac{\ell M}{2\rho }$ and $N=%
\frac{\gamma \ell M}{2\rho ^{2}}$, respectively. In the \textquotedblleft
light-cone\textquotedblright\ coordinates $u=\phi +\frac{\gamma \tau }{\ell }
$, $v=\phi -\frac{\gamma \tau }{\ell }$, the pseudosphere is locally
parameterized as \cite{Martinez},
\begin{equation}
\left(
\begin{array}{c}
x^{0} \\
x^{3}%
\end{array}%
\right) =\frac{\ell }{2\sqrt{2}}\left(
\begin{array}{cc}
\cos \alpha u & \sin \alpha u \\
-\sin \alpha u & \cos \alpha u%
\end{array}%
\right) \left(
\begin{array}{c}
(v+1)B+B^{-1} \\
(v-1)B-B^{-1}%
\end{array}%
\right) \,,  \label{x extreme}
\end{equation}%
with $(x^{1},x^{2})$ similarly obtained as the rotation by $\alpha u$ of $%
\frac{\ell }{2\sqrt{2}}((v-1)B+B^{-1},(v+1)B-B^{-1})$, for any $B$. Note
that the transformation now depends on the noncompact coordinate $v$, and%
\begin{equation}
ds^{2}=\ell ^{2}\left( \frac{dB^{2}}{B^{2}}-\alpha ^{2}du^{2}+\alpha
\,B^{2}\,dudv\right) \,.  \label{uv metric}
\end{equation}%
Since the metric does not depend on $u$, this coordinate can be compactified
and the product $\alpha u$\ now plays the role of a conical angle $0\leq
\alpha u=\phi _{c}\leq 2\pi (1-\alpha )$. The metric (\ref{uv metric}) is
equivalent to the one of the extremal $0$-brane for the particular choice
\begin{equation}
B(\rho )=\sqrt{\frac{1}{\alpha }\,\left( \frac{\rho ^{2}}{\ell ^{2}}+\alpha
^{2}\right) }\,,
\end{equation}%
and relating the defect to the negative mass, $\alpha =\sqrt{\frac{-M}{2}}$\
($M<0$). Here $u$\ is the only light-cone coordinate that is made periodic
upon identification $\alpha u=\phi _{c}~\simeq ~\phi _{c}+2\pi (1-\alpha )$,
whereas $v$\ is not identified and remains noncompact. Applying this
identification to (\ref{x extreme}), we find the Killing vector that
produces it,
\begin{equation}
\xi =\delta x^{A}\,\partial _{A}=2\pi \alpha \left( J_{03}-J_{12}\right)
=2\pi \alpha \,\partial _{\phi _{c}\,},
\end{equation}%
where we used that $\partial _{\phi _{c}}=\frac{\partial x^{A}}{\partial
\phi _{c}}\,\partial _{A}=J_{03}-J_{12}$. The source is therefore given by $%
j=2\pi \alpha \,\delta \left( T_{12}\right) \,\left( J_{03}-J_{12}\right) $,
as expected from the extremal limit.

\begin{figure}[tbp]
\includegraphics[width=0.5\textwidth]{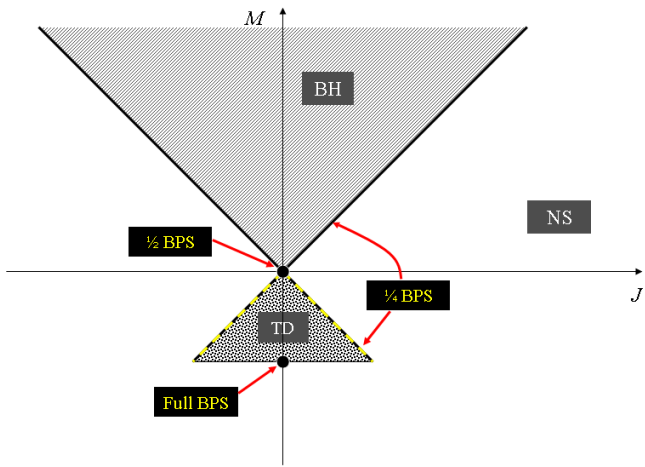}
\caption{Spectrum of the locally AdS 2+1 geometries ($\ell =1$). Region $%
M>|J|$ describes non-extremal BTZ black holes. The extremal configurations $%
M=|J|>0$, $M=J=0$ and $M+1=J=0$ are 1/4, 1/2, and full BPS states,
respectively. The configurations $M<-|J|$ are topological defects, and $%
M=-|J|$ are extremal 1/4 BPS.}
\label{BPSfig}
\end{figure}

The extremal $0$-branes coupled to the $2+1$ AdS supergravity admit globally
defined Killing spinors, defined by the condition $D\psi =0$. As shown in
Appendix \ref{BPS}, the spinning $0$-branes ($M<0$) are BPS states for the
extreme case only, $J=-\gamma \ell M$: for each sign of the spin ($\gamma
=\pm 1$), there is one globally defined Killing spinor given by
\begin{equation}
\psi =\sqrt{\frac{\rho }{\ell }-\frac{\ell M}{2\rho }}\,\left(
\begin{array}{c}
1 \\
-\gamma%
\end{array}%
\right) \,,  \label{killing spinor}
\end{equation}%
which is defined for $\rho >0$. Thus, apart from the well-known BPS states
provided by the AdS vacuum ($M=-1$), the extremal ($M\ell =|J|$), and the
massless 2+1 black holes \cite{Coussaert-Henneaux}, also the extremal $0$%
-branes are BPS states, which, like their positive mass counterparts, admit
one Killing spinor, that is, 1/4 of the supersymmetries admitted by the AdS
vacuum. This BPS solution seems to have been unnoticed in previous studies
\cite{Steif}. The different BPS states are depicted in Fig. \ref{BPSfig}.

\section{Summary and comments}

\textbf{1.} Generically, CS theories describe the dynamics of a (nonabelian)
connection in an odd-dimensional space, which can be viewed as the
worldvolume of a $2p$-brane. In fact, CS actions as well as their
interactions with external sources are structures of a similar nature \cite%
{Z,EZ}. The particular example of $0$-branes discussed here are topological
defects produced by an identification whichturn the 2+1 AdS manifold into an
orbifold with a naked singularity. It is shown that the resulting geometry
is a state in the negative part of the black hole's mass spectrum.

\textbf{2.} The lagrangian action for three-dimensional gravity with $%
\Lambda <0$ in the Chern-Simons representation is a metric-free, gauge
invariant and generally covariant object. The coupling between gravity and
an external source can be introduced through a minimal coupling of the form $%
\langle jA\rangle $, generalizing the one in electrodynamics, where the
2-form $j$ takes values in the gauge algebra. The resulting field equations
are $F=j$, where the current has the form $j\sim q\,\delta (T)d\Omega \,G$,
is exactly what is found for the static and spinning $0$-branes, produced by
angular defect, a\ naked singularity with \textquotedblleft negative
mass\textquotedblright . In the black hole case ($M>0$), there is no source
in the RHS of the field equations. In other words, black holes are not the
result of coupling gravity to a material source sitting somewhere in
spacetime; black holes are just exact solutions of the homogeneous
AdS-Einstein equations.

\textbf{3. }These NS define mathematically consistent couplings between $2+1$
gravity and external currents produced by point sources. Gauge invariance is
partially broken by the presence of a source like (\ref{spinning j}),
sitting at a fixed point in spacetime and pointing in a certain direction,
explicitly breaking AdS symmetry. Moreover, the current involves some
generators of the AdS group and therefore it is not invariant under the
action of the whole AdS gauge group, but only under a subgroup of it. Full
gauge invariance of the theory would be restored if the current $j$ becomes
dynamical, that is, if the current was produced by a dynamical field which
is also varied in the action, on the same footing as $A$.

\textbf{4.} The fact that these $0$-branes are NSs is not necessarily
inconsistent. Moreover, it was shown that, if endowed with the right amount
of angular momentum, they can be stable soliton-like objects that saturate
the Bogomolny'i bound. The fact that these supersymmetric states have
negative mass seems to contradict the common wisdom that supersymmetry
implies positivity of the energy spectrum. The point is that these are
supersymmetric extensions of the AdS --and not of the Poincar\'{e}-- group.
Indeed, the AdS vacuum has negative energy ($M=-1$) and it is perfectly
supersymmetric.

\textbf{5.} The presence of $\delta $-like distributions in the spacetime
manifold could be troublesome in General Relativity, especially if they
appear in the metric. In general, this is a problem because Einstein's
equations involve products and inverses of metric components, and such
operations are generically ill-defined in distribution theory. However,
writing ($2+1$)-dimensional gravity as a Chern-Simons theory circumvents
this problem because all field equations only involve exterior products of
forms (and no Hodge $\ast $-duals), which always give rise to sensible
distributional products.

\textbf{6.} A generalization to higher dimensions can proceed in two
different directions. One is to introduce a spherically symmetric
topological defect in a $S^{D-2}$, that is \emph{not} obtained by
identification with a Killing vector (except for $D=3$). In this way, $0$%
-branes in higher dimensional spaces are produced. Alternatively, if a
Killing vector is used to produce identifications in a two-dimensional
Euclidean plane in $D\geq 5$, introducing an angular deficit in $S^{1}$
only, (spinning) codimension 2 branes are obtained. These directions are
discussed in the extended paper \cite{Miskovic-Zanelli09}.

\textbf{7.} The topological defects discussed here could be the result of an
inhomogeneous collapse, as seen in $3+1$\ dimensions \cite{NumericalExp}.
Since non-extremal defects are non-BPS, they are probably unstable.
Non-extremal black holes can decay through Hawking radiation, but no
continuous decay mechanism is readily available for the topological defects
and they may disappear through a violent explosion. It could also be the
case that the states in the regions $J>|M|\ell $, $J<-|M|\ell $, and $M<-1$,
are forbidden by some general principle, and they might not form at all.

\section*{\textbf{Acknowledgments}}

The authors are grateful to C. Bunster, J. Edelstein, P. Mora, R. Olea and
S. Willison for enlightening conversations and suggestions, and to C. Mart%
\'{\i}nez for several useful comments and important discussions. This work
was supported in part by FONDECYT Grants 11070146, 1061291, 7070117 and
7080201. The Centro de Estudios Cient\'{\i}ficos (CECS) is funded by the
Chilean Government through the Millennium Science Initiative and the Centers
of Excellence Base Financing Program of Conicyt. CECS is also supported by a
group of private companies which at present includes Antofagasta Minerals,
Arauco, Empresas CMPC, Indura, Naviera Ultrags and Telef\'{o}nica del Sur.

\appendix{}

\section{BPS 0-branes in 3D CS-AdS supergravity \label{BPS}}

BPS states are bosonic solutions of the field equations that are invariant
under globally defined supersymmetry transformations. Here we show that the
extremal $0$-brane is a BPS state. The analysis carries over to other
examples in higher dimensions, with the appropriate supergoup in each case,
but the arguments are essentially the same. The reason for this is that all
CS supergravities \cite{CS-gravity-3D,CS-gravity} are gauge theories for an
essentially unique superalgebra that extends the AdS algebra in every
dimension. Thus, in all cases, the supersymmetry transformation of the
fermion (gravitini) takes the form
\begin{equation*}
\delta \psi =D\epsilon \text{ ,}
\end{equation*}%
where $D$ is the covariant exterior derivative for the connection
corresponding the the dimension in each case.

For $3$-dimensional Chern-Simons gravity, the minimal supersymmetric
extension of the AdS group is $OSp(2|1)$, and the connection (gauge field)
is
\begin{equation}
A=\frac{1}{2}\,\omega ^{ab}J_{ab}+\frac{1}{\ell }\,e^{a}J_{a}+\psi Q\,,
\end{equation}%
where $J_{AB}=\left\{ J_{a}:=J_{a3},\,J_{ab}\right\} $ are the AdS
generators satisfying $\left[ J_{AB},J_{CD}\right] =\eta _{AD}\,J_{BC}-\eta
_{BD}\,J_{AC}-\eta _{AC}\,J_{BD}+\eta _{BC}\,J_{AD}$ with $\eta _{AB}=\left(
-,+,+,+,-\right) $, and $Q^{\alpha }$ are the SUSY generators. The SUSY
transformation of the gravitino then takes the form%
\begin{equation}
\delta \psi =D\epsilon \equiv \left( d+\frac{1}{2}\,\omega ^{ab}\,J_{ab}+%
\frac{1}{\ell }\,e^{a}\,J_{a}\right) \epsilon \,.
\end{equation}%
The spinorial representation of the AdS generators in terms of $\gamma $%
-matrices that satisfy the Clifford algebra $\left\{ \gamma _{a}\,,\gamma
_{b}\right\} =2\eta _{ab}$, is%
\begin{eqnarray}
J_{a} &=&\frac{c}{2}\,\gamma _{a}\,, \\
J_{ab} &=&\frac{1}{4}\,\left[ \gamma _{a},\gamma _{b}\right] =\frac{1}{2}%
\,\varepsilon _{abc}\,\gamma ^{c}\,,
\end{eqnarray}%
where the constant $c=\pm 1$ corresponds to two inequivalent irreducible
representation of $\gamma $-matrices. We also use the identity $\left[
\gamma _{a},\gamma _{b}\right] =2\,\varepsilon _{abc}\gamma ^{c}$ (here $%
\varepsilon ^{012}=+1$). The Killing spinor equation then becomes%
\begin{equation}
D\epsilon =\left( d+\frac{1}{4}\,\varepsilon _{abc}\,\omega ^{ab}\gamma ^{c}+%
\frac{c}{2\ell }\,e^{a}\,\gamma _{a}\right) \epsilon =0\,.
\label{Killing spinor eq}
\end{equation}

We are interested in the spinning $0$-brane solutions ($-1<M<0$, $J\neq 0$)
defined throughout spacetime surrounding the $0$-brane ($\rho \neq 0$). The
non-vanishing components of the vielbein $e^{a}$ and spin-connection $\omega
^{ab}$ are
\begin{equation}
\begin{array}{ll}
e^{0}=f\,d\tau \,, & \omega ^{01}=\dfrac{\rho }{\ell ^{2}}\,d\tau +\rho
N\,d\phi \,,\medskip \\
e^{1}=\dfrac{1}{f}\,d\rho \,, & \omega ^{02}=\dfrac{N}{f}\,d\rho \,,\medskip
\\
e^{2}=\rho \left( Nd\tau +d\phi \right) \,,\qquad & \omega ^{12}=-f\,d\phi
\,,%
\end{array}%
\end{equation}%
where $f^{2}=-M+\frac{\rho ^{2}}{\ell ^{2}}+\frac{J^{2}}{4\rho ^{2}}$ and $%
N=-\frac{J}{2\rho ^{2}}$.

The radial component of the Killing spinor equation has the form%
\begin{equation}
D_{\rho }\epsilon =\left[ \partial _{\rho }+\dfrac{1}{2f}\,\left( N+\frac{c}{%
\ell }\right) \,\gamma _{1}\right] \epsilon =0\,,
\end{equation}%
and its general solution is%
\begin{equation}
\epsilon =\mathcal{M}(\rho )\,\varphi \,,
\end{equation}%
where $\varphi \,(\tau ,\phi )\,$is an arbitrary spinor and $\mathcal{M}%
(\rho )$ is the invertible matrix%
\begin{eqnarray}
\mathcal{M} &=&e^{-\gamma _{1}\,\eta }=\cosh \eta -\gamma _{1}\,\sinh \eta
\,, \\
\eta (\rho ) &\equiv &\int^{\rho }\dfrac{d\rho ^{\prime }}{2f(\rho ^{\prime
})}\,\left( N(\rho ^{\prime })+\frac{c}{\ell }\right) \,.
\end{eqnarray}%
The other two components of the Killing equation are%
\begin{eqnarray}
D_{\tau }\epsilon &=&\left[ \partial _{\tau }+\frac{1}{2\ell }\,U(\rho )%
\right] \epsilon =0  \label{DU} \\
D_{\phi }\epsilon &=&\left[ \partial _{\phi }-\frac{c}{2}\,U(\rho )\right]
\epsilon =0\,,  \label{DV}
\end{eqnarray}%
where we introduced the matrix%
\begin{equation}
U=c\,f\,\gamma _{0}+\dfrac{\rho }{\ell }\,\left( c\ell N-1\right) \,\gamma
_{2}\,.  \label{U}
\end{equation}%
In terms of $\varphi =\mathcal{M}^{-1}\epsilon $ and introducing the
light-cone coordinates%
\begin{equation}
u=\phi +\frac{c\tau }{\ell }\,,\qquad v=\phi -\frac{c\tau }{\ell }\,,
\end{equation}%
we have $D_{u}=\partial _{u}$ and $D_{v}=\partial _{v}-\frac{c}{2}\,U$, and
these differential equations can be written as%
\begin{equation}
\partial _{u}\varphi =0\,,\qquad \left( \partial _{v}-\frac{c}{2}\,\mathcal{M%
}^{-1}U\,\mathcal{M}\right) \varphi =0\,.  \label{spinor null directions}
\end{equation}

It is clear that these equations will have non-trivial solutions
in $\varphi (v)$ if and only if $\mathcal{M}^{-1}U\,\mathcal{M}$
is independent of $\rho $. Moreover, the spinor $\varphi $ must be
a null eigenvector of $\mathcal{M}^{-1}U\,\mathcal{M}$,
otherwise the solution of (\ref{spinor null directions}) could not
be single-valued: a nonzero eigenvalue would not be
periodic in the angle $\phi $. The representation of $\gamma $%
-matrices is $\gamma _{0}=-i\sigma _{2}\,,$ $\gamma _{1}=\sigma _{1}\,$and $%
\gamma _{2}=\sigma _{3}$ and, therefore, the condition for the existence of
a null eigenvector is
\begin{equation}
\det (\mathcal{M}^{-1}U\,\mathcal{M})=\det U=-M-\frac{cJ}{\ell }=0\,.
\end{equation}%
Thus, we conclude that there is a non-trivial Killing spinor only if
\begin{equation}
J=-c\ell M\,,
\end{equation}%
that is, the solution is extremal, and in that case $\varphi $ is a constant
spinor. The explicit form of $\varphi \ $that satisfies $U\,\varphi =0$ is%
\begin{equation}
\epsilon =\left(
\begin{array}{c}
\epsilon _{1} \\
-c\,\epsilon _{1}%
\end{array}%
\right) \,.
\end{equation}%
The lapse and shift functions in the extremal case are given by
\begin{equation}
f=\frac{\rho }{\ell }-\frac{\ell M}{2\rho }\,,\qquad N=\frac{c\ell M}{2\rho
^{2}}\,,
\end{equation}%
from where we find that the matrix $\mathcal{M}=e^{-\gamma _{1}\,\eta }$ is
determined by%
\begin{equation}
\eta =\ln f^{\frac{c}{2}}\,.
\end{equation}%
The condition that the spinor $\varphi =\mathcal{M}^{-1}\epsilon $ be
constant gives $\epsilon _{1}=\sqrt{f}$. Therefore, for each given sign of $J
$ (that fixes the representation $c$), there is one Killing spinor%
\begin{equation}
\epsilon =\sqrt{f}\,\left(
\begin{array}{c}
1 \\
-c%
\end{array}%
\right) \,,
\end{equation}%
that asymptotically behaves as the Killing spinor for zero-mass BTZ black
hole, $\left( \frac{\rho }{\ell }\right) ^{1/2}$ \cite{Coussaert-Henneaux}.


\begin{thebibliography}{99}
\bibitem{BTZ} M. Ba\~{n}ados, C.Teitelboim and J. Zanelli, \textit{The Black
hole in three-dimensional space-time, }Phys. Rev. Lett. \textbf{69}, 1849
(1992) [arXiv: hep-th/9204099].

\bibitem{Carlip} S. Carlip, \textit{The (2+1)-Dimensional black hole,}
Class. Quant. Grav. \textbf{12}, 2853 (1995); \textit{Gravity in
(2+1)-Dimensions}, Cambridge Univ. Pr. (1998).

\bibitem{Earman} It has even been suggested that green slime, lost socks and
TV sets could emerge from NS. See J. Earman, \textit{Bangs, crunches,
whimpers, and shrieks: Singularities and acausalities in relativistic
space-times,} Oxford Univ. Pr., NY, 1995.

\bibitem{NumericalExp} D. M. Eardley and L. Smarr, \textit{Time function in
numerical relativity. Marginally bound dust collapse},\ Phys. Rev. \textbf{D19}, 2239
(1979); D. Christodoulou, \textit{Violation of cosmic censorship in the
gravitational collapse of a dust cloud}, Commun. Math. Phys. \textbf{93},
171 (1984).

\bibitem{Jackiw90} R. Jackiw, \textit{Five lectures on planar gravity},
Lectures given at SILARG VII, Cocoyoc, Mexico, Dec 1990.

\bibitem{Miskovic-Zanelli09} O. Mi\v{s}kovi\'{c} and J. Zanelli, \textit{%
Couplings between Chern-Simons gravities and 2p-branes,} arXiv: 0901.0737
[hep-th].

\bibitem{BHTZ} M. Ba\~{n}ados, M. Henneaux, C. Teitelboim and J. Zanelli,
\textit{Geometry of the (2+1) black hole}, Phys. Rev. \textbf{D48} (1993)
1506.

\bibitem{Freedom} The parameters $2\pi a$ and $2\pi b$ are fixed modulo $%
2\pi $. We choose them so that the AdS space is recovered for $\phi
_{12}\simeq \phi _{12}+2\pi \,$and no identification in $\phi _{03}$.

\bibitem{Martinez} C. Mart\'{\i}nez, Private notes about the Killing spinors
for the BTZ black hole with $M>0$.

\bibitem{Coussaert-Henneaux} O. Coussaert and M. Henneaux, \textit{%
Supersymmetry of the (2+1) black holes}, Phys. Rev. Lett. \textbf{72} (1994)
183.

\bibitem{Steif} A. R. Steif, \textit{Supergeometry of three-dimensional
black holes}, Phys. Rev. \textbf{D53} (1996) 5521.

\bibitem{Z} J. Zanelli, \textit{Uses of Chern-Simons actions}, Ten Years of
AdS/CFT: AIP Conf. Proc. \textbf{1031}, 15 (2008).

\bibitem{EZ} J. Edelstein and J. Zanelli, \textit{Sources for Chern-Simons
theories}, arXiv: 0807.4217 [hep-th].

\bibitem{CS-gravity-3D} A. Ach\'{u}carro and P. K. Townsend, \textit{A
Chern-Simons Action for Three-Dimensional anti-De Sitter Supergravity
Theories}, Phys. Lett. \textbf{B 180} (1986) 89; E. Witten, \textit{%
(2+1)-Dimensional Gravity as an Exactly Soluble System}, Nucl. Phys. \textbf{%
B311} (1988) 46.

\bibitem{CS-gravity} A. H. Chamseddine,  \textit{Topological
gravity and supergravity in various dimensions}, Nucl. Phys. \textbf{B346} (1990) 213;
R. Troncoso and J. Zanelli, \textit{New gauge supergravity in
seven-dimensions and eleven-dimensions}, Phys. Rev. \textbf{D 58}: 101703
(1998); M. Ba\~{n}ados, R. Troncoso, and J. Zanelli, \textit{Higher
dimensional Chern-Simons supergravity}, Phys. Rev. \textbf{D 54} (1996) 2605.
\end{thebibliography}
\end{document}